\documentclass[preprint]{aastex}
\usepackage{emulateapj5}

\newcommand{\mbf}[1]{\mbox{\boldmath $#1$}}

\newcommand{\eqn}[1]{equation~(\ref{eqn:#1})}

\newcommand{\Sec}[1]{\S\ref{sec:#1}}

\newcommand{\Fig}[1]{Figure~\ref{fig:#1}}

\newcommand{\Tab}[1]{Table~\ref{tab:#1}}

\newcommand{\App}[1]{Appendix~\ref{app:#1}}

\newcommand{\im}{{\rm Im}}

\newcommand{\pauli}[1]{\ensuremath{\mbf{\sigma}_{#1}}}

\newcommand{\psr}{PSR\,J0437$-$4715}
\newcommand{\psrtwo}{PSR\,J1022$+$1001}

\newcommand{\Sinv}{S}
\newcommand{\norm}[1]{\|{#1}\|}

\shorttitle   {High-fidelity polarimetry and high-precision timing}
\shortauthors {W. van Straten}

\received{}

\begin{document}

\title{ High-Fidelity Radio Astronomical Polarimetry \\
  Using a Millisecond Pulsar as a Polarized Reference Source}

\author{W. van Straten}

\affil{Centre for Astrophysics and Supercomputing,
	Swinburne University of Technology, \\
	Hawthorn, VIC 3122, Australia}

\email{vanstraten.willem@gmail.com}

\begin{abstract}

  A new method of polarimetric calibration is presented in which the
  instrumental response is derived from regular observations of \psr\
  based on the assumption that the mean polarized emission from this
  millisecond pulsar remains constant over time.
  The technique is applicable to any experiment in which high-fidelity
  polarimetry is required over long time scales;
  it is demonstrated by calibrating 7.2~years of high-precision
  timing observations of \psrtwo\ made at the Parkes Observatory.
  Application of the new technique followed by arrival time estimation
  using matrix template matching yields post-fit residuals with an
  uncertainty-weighted standard deviation of 880~ns, two times smaller
  than that of arrival time residuals obtained via conventional
  methods of calibration and arrival time estimation.
  The precision achieved by this experiment yields the first
  significant measurements of the secular variation of the projected
  semi-major axis, the precession of periastron, and the Shapiro
  delay;
  it also places \psrtwo\ among the ten best pulsars regularly
  observed as part of the Parkes Pulsar Timing Array (PPTA) project.
  It is shown that the timing accuracy of a large fraction of the
  pulsars in the PPTA is currently limited by the systematic timing
  error due to instrumental polarization artifacts.
  More importantly, long-term variations of systematic error are
  correlated between different pulsars, which adversely affects the
  primary objectives of any pulsar timing array experiment.
  These limitations may be overcome by adopting the techniques
  presented in this work, which relax the demand for instrumental
  polarization purity and thereby have the potential to reduce the
  development cost of next-generation telescopes such as the Square
  Kilometre Array.

\end{abstract}



\section {Introduction}

High-precision pulsar timing exploits the exceptional 
rotational stability of millisecond pulsars to measure orbital
dynamics in binary systems and perform unique tests of general
relativity in the strong field regime \cite[e.g.][]{sta06}.
The basic measurement in timing analyses is the pulse time-of-arrival
(TOA), the epoch at which a fiducial phase of the pulsar's periodic
signal is received at the observatory.
The difference between the measured TOA and the arrival time predicted
by a best-fit model is known as the timing residual.
Timing residuals with a standard deviation of the order of 100\,ns
have been achieved for a growing number of millisecond pulsars
\citep[e.g.][]{lom01,hot07},
which has renewed both theoretical and experimental interest in the
detection of gravitational radiation using a pulsar timing array
\citep[PTA; e.g.][]{fb90,jb03,wl03,jhv+06,svc08,ych+11,vlj+11}.

The sensitivity of a PTA and the confidence limits
that may be placed on any gravitational wave detection directly depend
upon the precision and accuracy with which pulse arrival times can be
estimated.
Therefore, an important part of the ambitious PTA effort is the
quantification and reduction of systematic error through the
development of improved methods of arrival time estimation
\citep[e.g.,][]{hbo05,van06},
instrumental calibration \citep{ja98,van04c}, 
radio-frequency interference (RFI) mitigation \citep{kbr10,ng10},
compensation for the effects of propagation through the interstellar
medium \citep[e.g.,][]{yhc+07,hs08,crg+10},
statistical analysis of the stochastic nature of the pulsar signal
\citep{mth75,cd85,rr95,ovh+11},
and the characterization of timing noise \citep{cd85,sc10}.

At the Parkes Observatory, the majority of pulsar timing data are
obtained from observations of \psr, the closest and brightest
millisecond pulsar known.  Discovered in the Parkes 70-cm survey
\citep{jlh+93}, \psr\ has a spin period of $\sim 5.7$~ms; at 20~cm, it
has a pulse width at half maximum of about 130~$\mu$s \citep{nms+97}
and an average flux of 140~mJy \citep{kxl+98}.  Owing to its sharp
peak and large flux density, it is an excellent target for
high-precision pulsar timing studies.
It is also fortuitously located away from the Galactic plane
($l=245\degr$, $b=-42\degr$), which permits long observing sessions
during periods of local sidereal time when competition for the
telescope is relatively low.

\begin{figure*}
\centerline{\includegraphics[angle=-90,width=86mm]{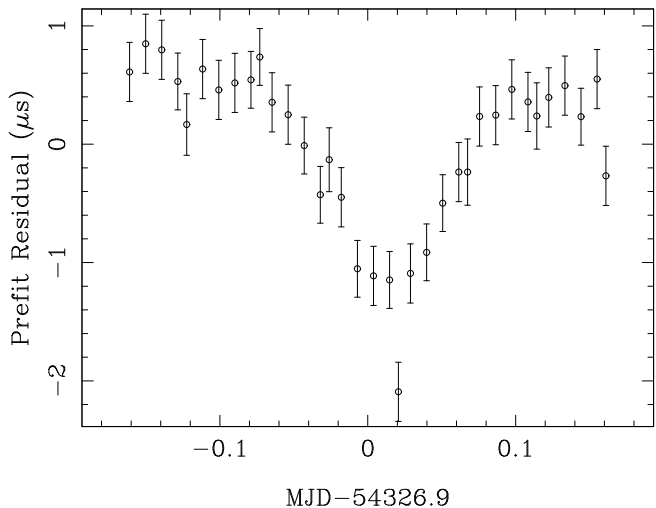}}
\caption{\label{fig:uncalibrated} Systematic timing error in one day
  of uncalibrated \psr\ observations.  As a function of time, the
  receiver rotates with respect to the sky, thereby geometrically
  altering the instrumental response to the highly polarized pulsar
  signal. The induced systematic error is about an order of magnitude
  larger than the estimated arrival time uncertainty ($\sim 250$\,ns).
}
\end{figure*}

Early timing of this pulsar highlighted the need for more accurate
polarimetric calibration \citep{sbm+97}.  The systematic timing error
due to instrumental polarization artifacts is readily observed in the dramatic
variation of arrival time residuals as a function of parallactic angle
(see \Fig{uncalibrated}). 
This issue motivated the discovery and development
of the polarimetric invariant profile \citep{bri00}.
Arrival times derived from the invariant profile proved to be
significantly more accurate than those of the total intensity profile,
a breakthrough that led to the first detection of annual-orbital parallax
and a new test of general relativity \citep{vbb+01}.

Despite its demonstrated advantages over the total intensity profile,
there remain a number of drawbacks to the use of the invariant
profile.  First, the signal-to-noise ratio $S/N$ of the invariant
profile is lower than that of the total intensity, reaching zero for
completely polarized sources.  Second, the noise in the invariant
profile is neither normally distributed nor homoscedastic with respect to
pulse longitude (see \App{invariant}); these two properties are
assumed, either explicitly or implicitly, by most template matching
algorithms commonly used for pulsar timing.  Finally, the computation
of the invariant profile requires estimation and correction of the
bias due to noise, and the relative precision of the bias estimate is
inversely proportional to the square root of the number of discrete
samples within off-pulse regions of pulse longitude.  For a typical
observation of \psr, the relative error in the bias correction is of
the order of 8\%, an unacceptably large uncertainty when aiming for
arrival time uncertainty that is four orders of magnitude smaller than
the pulse width.

These concerns motivated the development of matrix template matching
\citep[MTM;][]{van06}, a technique that exploits the additional timing
information in the polarized component of the pulsar signal while
simultaneously eliminating residual polarimetric calibration errors.
For some pulsars, the mean polarization profile exhibits sharper
features than observed in the total intensity;
%
%
these features generate more power at higher harmonics of the pulsar
spin frequency, yielding greater arrival time precision than might be
expected from the polarized flux alone \citep[e.g. see Figure 1 of][]{van06}.
%
%
Furthermore, MTM improves arrival time accuracy by mitigating
systematic timing errors due to instrumental polarization artifacts.

Although the development of MTM was primarily driven by the
high-precision timing of \psr, \citet{van06} predicted that \psrtwo\ would
benefit most from this technique, including a $\sim 33\%$ increase in
timing precision and a significant decrease in systematic timing error.
\psrtwo\ was contemporaneously discovered in both the Princeton-Arecibo
Declination-Strip Survey \citep{cnst96} and the Green Bank Northern
Sky Survey \citep{snt97}.
\citet{cnst96} observed variations in the shape of its mean pulse
profile on a timescale of the order of minutes and noted that the
accuracy of arrival time estimates was limited by this effect.
Mean profile variations over such timescales would be distinct from
pulse-to-pulse shape variations such as giant pulses, microstructure,
and drifting sub-pulses \citep[e.g.][]{han71,lsg71,bac73}, which when
averaged over many pulses can be described as phase jitter
\citep{cd85} or stochastic wide-band impulse-modulated self-noise
\citep[SWIMS;][]{ovh+11}.
As it is typically assumed that the mean pulse profile does not change with
time, detection of systematic profile variations in a millisecond pulsar
would have a serious impact on the field of high-precision timing.

Accordingly, the stability of the \psrtwo\ pulse profile has been
the subject of a number of detailed studies \citep{kxc+99,rk03,hbo04}.
These are reviewed in \Sec{systematic}, following a mathematical
treatment of the systematic timing error due to instrumental
distortion of the total intensity profile.
\Sec{observation} describes the observing system and the novel
procedures used to calibrate the \psrtwo\ data for this study,
including a new scattered power correction algorithm that is described
in more detail in \App{spc}.
As part of this analysis, a 7.2-year history of the polarimetric
response at Parkes is presented and its impact on systematic timing
error is discussed.
In \Sec{discussion}, the results of this study are compared and
contrasted with previous work, the relevance of the calibration method
to other pulsars is discussed, and potential future directions are
considered.


\section{Systematic Timing Error}
\label{sec:systematic}

In the following discussion, the polarization of electromagnetic
radiation is described by the second-order statistics of the
transverse electric field vector $\mbf{e}$, as represented by the
four Stokes parameters
\begin{equation}
S_k = \langle \mbf{e}^\dagger \pauli{k} \mbf{e} \rangle.
\end{equation}
The angular brackets denote an ensemble average; $\mbf{e}^\dagger$ is
the conjugate transpose of $\mbf{e}$; $\pauli{0}$ is the $2\times2$
identity matrix; \pauli{1}, \pauli{2}, and \pauli{3} are the Pauli
spin matrices; $S_0$ is the total intensity, and $\mbf{S} =
(S_1,S_2,S_3)$ is the polarization vector \citep[e.g.][]{bri00}.
The reception and propagation of polarized radiation is described by
linear transformations of $\mbf{e}$, as represented using complex
$2\times2$ Jones matrices.
Any non-singular matrix can be decomposed into the product of a
unitary matrix and a self-adjoint matrix.
Unitary transformations result in a three-dimensional Euclidean
rotation of the Stokes polarization vector, leaving the total
intensity unchanged and preserving the degree of polarization.
Self-adjoint Jones matrices effect a Lorentz boost of the Stokes
4-vector, mixing total intensity and polarized flux \citep{bri00}.
The boost component, also known as the {\it poldistortion}
\citep{ham00}, of the instrumental response distorts the shape of
the total intensity pulse profile and introduces systematic arrival
time errors.
Physically, instrumental boost transformations are caused by
differential amplification of the signals from the two
orthogonally-polarized receptors and by cross-coupling between these
receptors.

In \cite{van06}, the systematic timing error induced by a given level
of polarization distortion is predicted for a number of millisecond
pulsars using a numerical simulation in which the template profile for
each pulsar is copied and subjected to a boost transformation.
The orientation of the boost axis in Poincar\'e space is varied, and
the minimum and maximum induced phase shifts between the distorted
total intensity profile and the template profile are recorded.
It is also possible to predict the systematic timing error due to
polarization calibration errors by computing the rate at which the
best estimate of the phase shift varies with respect to the
instrumental boost transformation parameters.
Beginning with equation (12) of \cite{van06}, the best estimate of the
phase shift $\varphi$ derived using only the total intensity is given
by
\begin{equation}
\varphi = {
{ \sum \phi_{0,m} |S_{0,m}| \nu_m }
\over
{ 2\pi \sum |S_{0,m}| \nu_m^2 }
}.
\label{eqn:varphi_S0}
\end{equation}
Here, $S_{0,m}=S_0^{\prime*}(\nu_m)S_0(\nu_m)$ is the
cross-spectral power of the observed total intensity $S_0^{\prime}$
and the template total intensity $S_0$ in the Fourier
domain, $\phi_{0,m}$ is the argument of $S_{0,m}$, and the summations
in the numerator and denominator are performed over $m=1$ to $m=N$,
where $N$ is the maximum harmonic at which the fluctuation power
spectrum exhibits significant power.

Now consider an observed total intensity profile that is a copy of the
template subjected to an instrumental boost transformation, as
parameterized in Section 4.1 of \cite{van04c} such that
$\sinh^2\beta=\mbf{b\cdot b}$.  Referring to equation (10) of
\cite{bri00}, the transformed total intensity is given by
\begin{equation}
S_0^\prime = (1+2\,\mbf{b\cdot b}) S_0 
	+ 2 b_0 \, \mbf{b\cdot S},
\end{equation}
where $b_0 = \cosh\beta = \sqrt{1+\mbf{b\cdot b}}$.
The partial derivatives of $S_0^\prime$ with respect to the boost
parameters $b_k$ are used to compute the
partial derivatives of the phase shift with respect to these parameters
when the boost parameter $\beta$ is zero,
\begin{equation}
\dot\varphi_k \equiv {\partial \varphi \over \partial b_k }\bigg|_{\beta=0}
= {{ \sum \im [S_k^*(\nu_m) S_0(\nu_m)] \nu_m }
   \over
   { \pi \sum |S_0(\nu_m)|^2 \nu_m^2 }}.
\label{eqn:susceptibility}
\end{equation}
The above expression defines the components
of a three-dimensional gradient $\mbf{\dot\varphi}$. 
The magnitude of the gradient
%
$\dot\varphi_\beta \equiv |\mbf{\dot\varphi}|$
%
provides a measure of the susceptibility of arrival time estimates
from a given pulsar to instrumental distortion.

To aid in the physical interpretation of the susceptibility
$\dot\varphi_\beta$, the first order approximations to equations (25)
and (32) of \cite{van06} are used to define the systematic timing
error induced by either a 1\% differential gain error or receptor
non-orthogonality of 0.01 radians ($\sim$ 0.6\degr),
\begin{equation}
\tau_\beta \equiv 5\times10^{-3} \dot\varphi_\beta P,
\end{equation}
where $P$ is the pulsar spin period.
Values of $\tau_\beta$ for each of the 20 millisecond pulsars that are
regularly observed as part of the Parkes Pulsar Timing Array
\citep[PPTA;][]{mhb+12} project are reported in \Tab{theory}.
%
%
\begin{table*}
\caption{Relative Arrival Time Uncertainties and Systematic Timing Error} 
\begin{center} 
\begin{tabular}{llll} 
\tableline 
\tableline 
Pulsar & $\hat\sigma_{\varphi}$ & $\hat\sigma_{\tilde\varphi}$ & $\tau_\beta$ (ns)\\ 
\tableline 
J0437$-$4715 & 0.85 & 1.43 & 207 \\
J0613$-$0200 & 0.92 & 1.46 & 59 \\
J0711$-$6830 & 0.88 & 1.54 & 81 \\
J1022+1001 & 0.68 & 1.65 & 282 \\
J1024$-$0719 & 0.74 & 2.11 & 34 \\
J1045$-$4509 & 0.88 & 1.48 & 338 \\
J1600$-$3053 & 0.90 & 1.39 & 115 \\
J1603$-$7202 & 0.85 & 1.55 & 142 \\
J1643$-$1224 & 0.91 & 1.40 & 266 \\
J1713+0747 & 0.85 & 1.58 & 6 \\
J1730$-$2304 & 0.71 & 1.70 & 198 \\
J1732$-$5049 & 0.96 & 1.38 & 185 \\
J1744$-$1134 & 1.56 & 6.43 & 105 \\
J1824$-$2452A & 0.88 & 2.56 & 18 \\
J1857+0943 & 0.89 & 1.43 & 124 \\
J1909$-$3744 & 1.02 & 1.51 & 22 \\
J1939+2134 & 0.95 & 1.49 & 44 \\
J2124$-$3358 & 0.85 & 1.45 & 127 \\
J2129$-$5721 & 1.15 & 1.61 & 211 \\
J2145$-$0750 & 0.95 & 1.44 & 147 \\
\tableline 
\end{tabular} 
\end{center} 
\label{tab:theory}
\end{table*}
The quantities in this table are derived from the high $S/N$
polarization profiles presented by \citet{ymv+11}.
For the majority of pulsars in the PPTA, a modest degree of
instrumental distortion induces systematic timing error of the order
of 100~ns, which is sufficient to seriously impede progress toward the
detection of the stochastic gravitational wave background
\citep{jhlm05}.

To first order, the systematic timing error due to instrumental
distortion is given by the inner product, $\mbf{b \cdot \dot\varphi}$.
In the polar coordinate system best suited to describing linearly
polarized receptors,
$\mbf{b}\simeq(\gamma,\delta_\theta,\delta_\epsilon)$, where $\gamma$
parameterizes the differential receptor gain, and $\delta_\epsilon$
and $\delta_\theta$ describe the non-orthogonality of the receptors
\citep[see Section 3.3 of][]{van06}.
These properties of the instrument may vary as a function of time for
a variety of reasons.
Diurnal variations are introduced by the parallactic rotation of the
receiver with respect to the sky, as shown in \Fig{uncalibrated}.
Furthermore, step changes in differential gain are introduced when the
levels of attenuation applied to each of the two orthogonal
polarizations are reset at the start of each observation and/or
periodically updated during the observation.
In contrast, the cross-coupling of the receptors is typically assumed
to remain relatively stable over timescales of the order of months.

Of particular interest to a long-term timing experiment such as a
pulsar timing array are any step changes in $\mbf{b}$ due to
modifications of the instrumentation and/or slow variations in
$\mbf{b}$ due to gradual degradation of one or more system components.
The temporal variations in distortion-induced error due to the
long-term evolution of a given instrument will be correlated between
the different pulsars observed using that system.
In an array of $N$ pulsars, the spectral structure of the correlated
systematic error due to instrumental distortion is described by
\begin{equation}
\label{eqn:covariance}
{\bf C}_\tau(f) = {\mbf \Upsilon} {\bf C}_\beta(f) {\mbf \Upsilon}^T,
\end{equation}
where ${\bf C}_\tau$ is the $N\times N$ matrix of the auto- and
cross-power spectra of the timing residuals \citep[e.g.][]{ych+11},
$\mbf\Upsilon$ is the $N \times 3$ Jacobian matrix,
\begin{equation}
\label{eqn:jacobian}
\Upsilon_j^k \equiv {\partial\tau_j\over\partial b_k}
= P_j {\partial\varphi_j\over\partial b_k}
\end{equation}
($P_j$ is the spin period of the $j$th pulsar) and ${\bf C}_\beta(f)$
is the $3\times3$ matrix of the auto- and cross-power spectra of the
instrumental distortion parameters (the components of $\mbf b$).

In practice, an optimally-weighted sum of the cross-power spectral
harmonics is used to define a detection statistic with maximum
sensitivity to a stochastic background of low-frequency gravitational
waves \citep[e.g.][]{abc+09,ych+11}.
Given only the mean polarization profile of each pulsar in the timing
array, it is possible to predict the impact of correlated systematic
timing error due to instrumental distortion on such a detection
statistic.
First, only the dominant harmonic of the cross-power spectrum of two
pulsars with distortion gradients $\mbf{\dot\varphi}_A$ and
$\mbf{\dot\varphi}_B$ is considered.
Next, it is assumed that the instrumental distortion
parameters are uncorrelated and homoscedastic, such that ${\bf
  C}_\tau=\sigma_\tau^2{\bf I}$, where {\bf I} is the $3\times3$
identity matrix.
Following these simplifications, a first-order approximation of the
coefficient of correlated systematic timing error is given by
\begin{equation}
c_{AB} =
 { \mbf{\dot\varphi}_A\, \mbf{\cdot}\, \mbf{\dot\varphi}_B
\over
 |\mbf{\dot\varphi}_A| |\mbf{\dot\varphi}_B| }.
\label{eqn:cross}
\end{equation}
This correlation coefficient depends only on the shape of the mean
polarization profile and has no modal structure on the celestial
sphere.
That is, the correlated timing error due to instrumental distortion
will corrupt all of the moments in a multipole expansion of timing
delay, which will impact on all of the primary goals of any pulsar
timing array experiment \citep{fb90}, including
the long-term measurement of time \citep[monopole moment;][]{gp91},
corrections to the Solar System ephemeris \citep[dipole moment;][]{chm+10},
and detection of the gravitational wave background \citep[quadrupole
moment;][]{hd83}.

\begin{figure*}
\centerline{\includegraphics[angle=0,width=86mm]{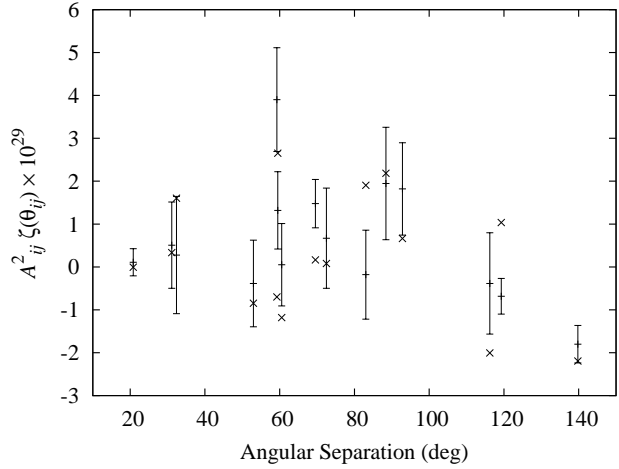}}
\caption{\label{fig:ych+11}
Comparison between predicted and measured estimates of correlation 
in pulsar timing residuals.
The predicted coefficients of correlated systematic timing error,
 $c_{AB}$ (crosses) are qualitatively similar to the optimally-weighted sum of 
cross-spectral power estimates $A_{ij}^2\zeta(\theta_{ij})$
(points with error bars) presented by \cite{ych+11}.
The values of $c_{AB}$ have been scaled by $2.7\times10^{-29}$, as
determined by a (non-weighted) least-squares fit to the values of
$A_{ij}^2\zeta(\theta_{ij})$.  }
\end{figure*}

To illustrate the potential importance of correlated systematic timing
error, the high $S/N$ polarization profiles presented by
\citet{ymv+11} are used to compute the cross-correlation coefficients
for each of the 15 pulsar pairs presented in Figure 4 of
\cite{ych+11}.
The cross-spectral power measurements of \citet{ych+11} are compared
with the predicted values of $c_{AB}$ in \Fig{ych+11}; the coefficient
of correlation between predicted and measured values is
$\rho\sim0.35$.
The agreement between predicted and experimental measures of
correlation indicates that the systematic error due to
instrumental distortion is a plausible contributor to
the apparent anti-correlation between the results of \cite{ych+11}
and the expected quadrupolar signature predicted by \cite{hd83}.

Also shown in \Tab{theory} are the relative arrival time uncertainties
$\hat\sigma_{\varphi}$ and $\hat\sigma_{\tilde\varphi}$ defined in
\citet{van06}.  The former is the ratio between the theoretical error
in arrival time estimates derived using matrix template matching (MTM)
and the error in estimates derived using only the total intensity; the
latter is the ratio between the uncertainties in arrival times derived
from the invariant interval and from the total intensity.
The statistical uncertainty in the invariant profile arrival times is
underestimated because the effects of heteroscedasticity and bias
correction error are not considered.
Regardless, for every pulsar in the PPTA, precision is lost through
use of the invariant interval.
The increase in uncertainty is most dramatic for PSR~J1744$-$1134, which
emits almost 100\% linearly polarized radiation.  For this pulsar, MTM
is also predicted to yield arrival times with greater error than those
derived from the total intensity owing to the large coefficient of
multiple correlation ($\sim 0.9$) between the phase shift used to
compute the TOA and the model parameters that describe the
polarization transformation.

\newpage

\subsection{PSR\,J1022+1001}
\label{sec:motivation}

\Tab{theory} also predicts that matrix template matching will yield the
greatest relative improvement in arrival time precision for \psrtwo.
This pulsar also has the second-highest value for predicted systematic
timing error owing to the susceptibility of the total intensity
profile to polarization distortion.
This may explain the profile shape variations and excess
arrival time error first noted by \citet{cnst96}.
In a detailed study of the spectrum and polarization of these
variations using the Effelsberg 100-m Radio Telescope,
\citet{kxc+99} assert that the observed fluctuations in pulse shape
cannot be explained by instrumental effects.
However, the data used in this analysis were calibrated using the
``ideal feed'' assumption that there is no cross-coupling between
receptors and that the reference source used for gain calibration
produces an equal and in-phase response in each receptor.
Given that receptor cross-coupling of the order of 10\% to 20\% is
commonly observed when more accurate calibration is performed
\citep[e.g.][]{scr+84,xil91,van04c}, an instrumental origin of the
variability deserves consideration.

\begin{figure*}
\centerline{\includegraphics[angle=-90,width=50mm]{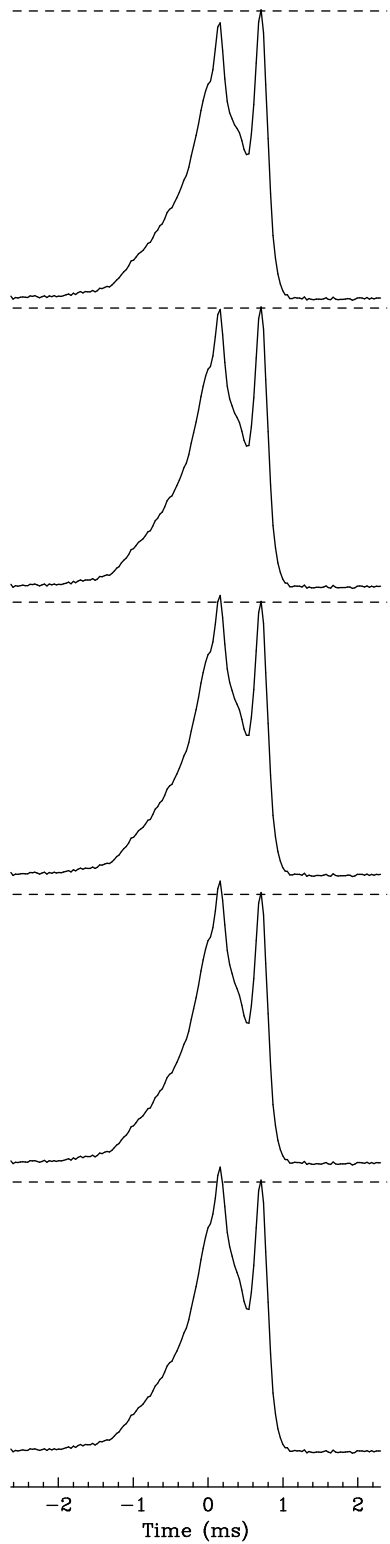}}
\caption {Simulated temporal variation of the \psrtwo\ mean total
  intensity profile.  The parallactic angle of each simulated
  observation is computed for the Effelsberg 100-m Radio Telescope at
  the epoch of the corresponding panel in Figure 2 of \citet{kxc+99}.
  Over this 3.3\,hr observation, the parallactic angle varies by
  37\degr. }
\label{fig:kxc+99}
\end{figure*}

For instance, the temporal variation of the total intensity profile
observed by \citet{kxc+99} is reproducible using a simple model in which
the polarized signal is rotated by the parallactic angle of the
receiver and then subjected to an instrumental boost that mixes $\sim$10\%
of the linearly-polarized flux with the total intensity.
The similarity between the simulated observations plotted in
\Fig{kxc+99} and the total intensity profiles presented in Figure 2 of
\cite{kxc+99} supports the argument that the variations are the
result of uncalibrated instrumental distortion.

Similarly, the small frequency-scale variations observed by
\citet{kxc+99} may be readily explained by spectral variation of the
instrumental response \citep[e.g.\ Figure 3 of][]{van04c}.
The observations used to demonstrate the narrow-band variations and 
presented in Figure 8 of \citet{kxc+99}
were recorded using the Effelsberg-Berkeley-Pulsar-Processor
with a bandwidth of 56~MHz.
In this mode, only the fluxes in each of the circularly polarized
signals (i.e.\ two out of four Stokes parameters) are retained;
consequently, only the mixing between circularly-polarized flux and
total intensity can be calibrated in these data.
Therefore, frequency-dependent mixing between linearly-polarized
flux and total intensity is a plausible explanation for the observed
profile shape variations.

\citet{rk03} analyzed the variability of \psrtwo\ using multi-frequency
observations made at the Westerbork Synthesis Radio Telescope (WSRT).
They argue that, because the WSRT is composed of equatorially-mounted
antennas, it is possible to rule out pulse shape variations due to
instrumental distortion combined with parallactic rotation of the feed.
However, the polarimetric response of a phased array can vary in both
time and frequency for a wide variety of reasons, including imperfect
phase calibration and (as for any single-dish antenna) instrumental gain
variations.
Indeed, the automatic gain controllers used to prevent instrumental
saturation \citep{vkv02} also make it impossible to accurately
calibrate the polarimetric response of the WSRT phased array
\citep{es04}.

The high degree of susceptibility of the \psrtwo\ pulse profile to
polarization distortion and the inadequate instrumental calibration
employed in these previous works cast doubt on the conclusion that the
profile variations are intrinsic to the pulsar.
Furthermore, \citet{hbo04} find no evidence of significant profile
shape variations in \psrtwo\ observations made at the Parkes
Observatory.
This study achieved arrival time residuals with a standard deviation
of the order of 2.3~$\mu$s and $\chi^2$ per degree of freedom $\sim
1.4$,
whereas \citet{kxc+99} obtained a reduced $\chi^2$ close to unity only
after systematic errors of the order of 10~$\mu$s were added in
quadrature to the formal uncertainties.

The analysis by \citet{hbo04} is based on observations integrated over
5~min and 64~MHz and spanning 1.3 years.
Increasing the integration length to 60~min and the time span to 2.3
years yielded arrival time residuals with a standard deviation of the
order of 1.5~$\mu$s and reduced $\chi^2\sim3$ \citep{hbo06}.
With the same instrumentation and similar integration lengths,
\cite{vbc+09} achieved 1.6~$\mu$s residuals in data spanning 5.1 years.
In the absence of systematic error and other sources of noise (e.g.
dispersion variations), the one-hour integrations used in these two
studies should have yielded timing residuals of the order of 660~ns.
This is of the same order as the systematic timing error
introduced by a differential gain error of $\sim2.3\%$ or by receptor
non-orthogonality of $\sim1.4\degr$ (cf. \Tab{theory}).
However, the arrival times used in these studies were derived from
either uncalibrated \citep{hbo06} or inaccurately calibrated
\citep{vbc+09} total intensity pulse profiles, which are highly
susceptible to instrumental distortion.
Therefore, it is reasonable to anticipate that high-fidelity
polarimetric calibration will improve the accuracy of \psrtwo\ arrival
time estimates.

This conjecture is verified in the following section, which describes a
new method of deriving the instrumental response by using \psr\ as a
polarized reference source.
The method is used to calibrate high-precision timing observations of
\psrtwo\ and demonstrated to reduce the systematic error due to
instrumental polarization distortion.


\section{Observations and Analysis}
\label{sec:observation}

Dual-polarization observations of \psr\ and \psrtwo\ were made at the
Parkes Observatory using the H-OH receiver and the center element of
the Parkes 21-cm Multibeam receiver \citep{swb+96}.
Two 64-MHz bands, centered at 1341 and 1405\,MHz, were digitized using
two bits per sample and processed using the second generation of the
Caltech-Parkes-Swinburne Recorder \citep[CPSR2;][]{bai03,hot07}.
To maintain optimal linear response during digitization, the detected
power was monitored and the sampling thresholds were updated
approximately every 30 seconds.
The baseband data were reduced using {\sc dspsr} \citep{vb11}, which
corrects quantization distortions to the voltage waveform using the
dynamic level setting technique \citep[hereafter JA98]{ja98}, then
performs phase-coherent dispersion removal \citep{han71,hr75} while
synthesizing a 128-channel filterbank \citep{van03a}.  The Stokes
parameters are then formed and integrated as a function of topocentric
pulse phase, producing uncalibrated polarization profiles with 1024
discrete pulse longitudes.

\subsection{Minimizing quantization distortions}
\label{sec:distortions}

During analog-to-digital conversion, the radio signal is subjected to
non-linear distortions that significantly alter the observed pulse
profile \citep[JA98;][]{kv01}.  JA98 introduce a dynamic output level
setting technique that is employed by {\sc dspsr} to correct the
underestimation of undigitized power.  To implement this
technique, the digitized data are divided into consecutive segments of
$L$ samples and, for each segment, the number of low-voltage states
$M$ is counted.  A histogram of occurrences of $M$ is archived with
the pulsar data and used offline to evaluate the quality of the
digital recording.

When the voltage input to the digitizer is normally distributed, the
ratio $\Phi=M/L$ has a binomial distribution as in equation (A6) of
JA98.  The difference between this theoretical expectation and the
recorded histogram of $M$ provides a measure of the degree to which
either the input signal is not well described by a normal
distribution or the sampling thresholds diverge from optimality (e.g.
at the start of an observation).  This difference, called the {\it
  two-bit distortion}, is given by
\begin{equation}
D = \sum_{M=0}^L \left[{\mathcal P}(M/L) - {\mathcal H}(M) \right]^2
\label{eqn:d2bit}
\end{equation}
where ${\mathcal P}(\Phi)$ is the expected binomial distribution and
${\mathcal H}(M)$ is the recorded distribution of $M$.
Separate histograms of $M$ are maintained for each polarisation,
and the reported distortion is simply the sum of the distortion
in each polarisation.

\begin{figure*}
\centerline{\includegraphics[angle=-90,width=86mm]{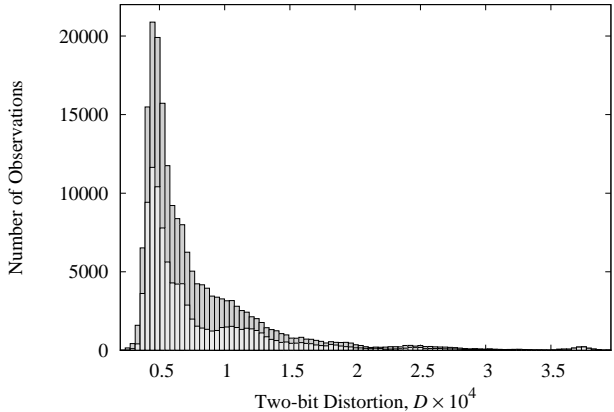}}
\caption{\label{fig:d2bit}
Stacked histograms of the two-bit distortion $D$ in  
the CPSR2 bands centered at 1341\,MHz (top/gray) and 1405\,MHz (bottom/light gray).}
\end{figure*}

For each 16.8-second integration output by CPSR2, the two-bit
distortion $D$ was computed and all data with $D > 3.5\times10^{-4}$
were discarded.  This threshold was chosen using the histograms plotted
in \Fig{d2bit};
beyond the range of plotted values, a sparse tail of
outliers extends to a maximum value of $D\sim0.98$.  Approximately
3.3\% of the observations were rejected; the remaining data were
corrected for scattered power using the method described in \App{spc}
and implemented in
the {\sc psrchive} software package for pulsar data archival and
analysis\footnote{\tt http://psrchive.sourceforge.net} \citep{hvm04}.
The corrected data were then summed until the integration length 
was of the order of five minutes.


\subsection{Polarimetric template profile}
\label{sec:template}

To use a pulsar as a polarized reference source requires a high
$S/N$, well-calibrated, template polarization profile.
From over 700 hours of \psr\ observations were selected
only those sessions with low levels of two-bit distortion and of
sufficient duration to achieve a precise estimate of the instrumental
response.
About 50 sessions, each $\sim$ 8 hours in duration and
with corresponding pulsed noise diode observations, were calibrated
using measurement equation modeling \citep[MEM;][]{van04c}.
The observed Stokes parameters were normalized to account for pulsar
flux variations as described in \App{invariant}.
As in \citet{van04c}, the two degenerate model parameters (a boost
along the Stokes $V$ axis and a rotation about this axis) were
constrained by assuming that observations of 3C 218 (Hydra~A) have
negligible circular polarization and that the orientation of the
receiver is known.
Although these assumptions are not necessarily valid, absolute
certainty in the Stokes parameters is not required for the purposes of
a high-precision timing experiment.
It is sufficient that the observed Stokes parameters can be mapped to
the template profile by a single Jones transformation.

This prerequisite may be evaluated using the objective merit function
of the MTM least-squares fit; the reduced $\chi^2$ statistic will be
greater than unity whenever the transformation between the template
and the observed pulse profile is not well-described by a single Jones
matrix.
Various phenomena may contribute to a poor model fit.
For example, if the mean polarized emission from the pulsar evolves
over time, then the difference between the template and the observed
profile will vary as a function of pulse longitude in a manner that
cannot be adequately modeled by a single Jones transformation.
Furthermore, as Jones matrices describe only linear transformations of
the electric field, matrix template matching will fail to model any
non-linear component of the instrument.

\begin{figure*}
\centerline{\includegraphics[angle=-90,width=86mm]{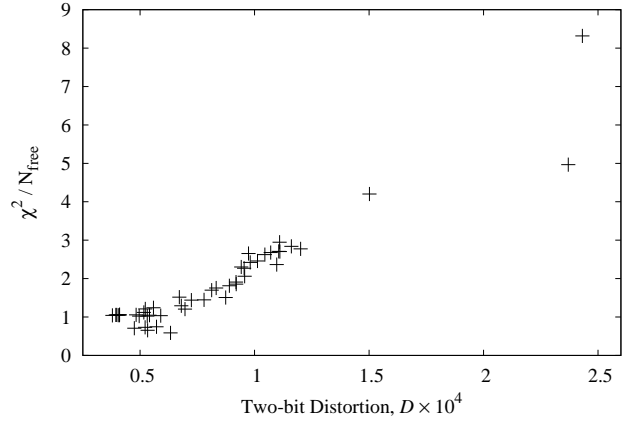}}
\caption { Average matrix template matching goodness-of-fit
  $\chi^2/N_\mathrm{free}$ as a function of median two-bit distortion
  $D$ in the CPSR2 band centered at 1341~MHz. }
\label{fig:chisq_d2bit}
\end{figure*}

Analog-to-digital conversion using only two bits per sample is an
intrinsically non-linear process;
therefore, variations in the response of the digitizer adversely
affect the MTM goodness-of-fit, as shown in \Fig{chisq_d2bit}.
Here, 5 of the $\sim50$ MEM-calibrated \psr\ observations were
integrated to produce a template profile and the mean polarization
profiles from each of the original $\sim50$ observations were fit to
this template using MTM.
The reduced $\chi^2$ of each MTM fit (averaged over all frequency
channels) is highly correlated with the median value of the two-bit
distortion $D$ in each session; i.e.\ greater distortion reduces the
merit of the MTM fits.
For the 5 observations used to create the template profile,
$\chi^2/N_\mathrm{free} < 1$ because covariance between the
observation and the template is not included in the definition of the
MTM objective merit function.

\Fig{chisq_d2bit} justifies the experimental design decision to
regularly update the two-bit sampling thresholds during each observation.
Independent sampling thresholds were applied to each of the two
orthogonal polarizations;
consequently, the astronomical signal was subjected to an unknown
differential gain that varies with time.
Such differential gain variations can be modeled using MTM,
whereas the non-linear response of the digitizer cannot.
Therefore, when recording the baseband signal with a two-bit
digitizer, it is more important to maintain optimal linear response
than it is to maintain a constant flux scale.

\begin{figure*}
\centerline{\includegraphics[angle=-90,width=86mm]{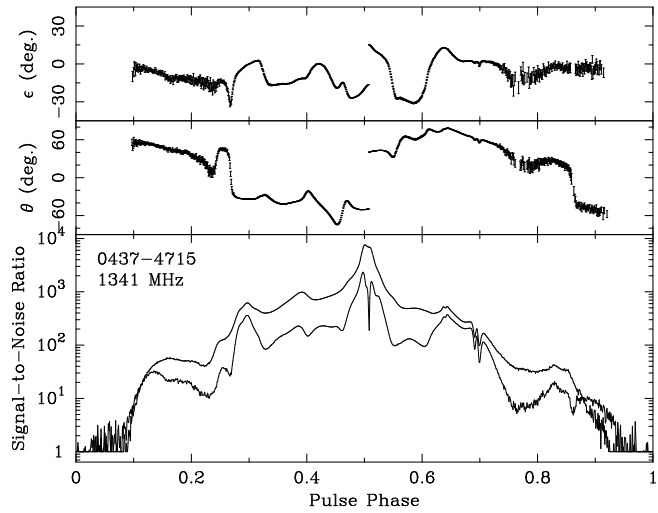}}
\caption {Mean polarization of \psr, plotted as a function of pulse
  phase using polar coordinates: ellipticity, $\epsilon$, orientation,
  $\theta$, and polarized intensity, $S=|\mbf{S}|$ (plotted in grey
  below the total intensity, $S_0$).  Flux densities are normalized by
  $\sigma_0$, the standard deviation of the off-pulse total intensity
  phase bins.  Data were integrated over a 64\,MHz band centered at
  1341\,MHz for approximately 96~hours. }
\label{fig:0437}
\end{figure*}

Out of the $\sim 50$ sessions that were calibrated using MEM, 16 were
selected with low median two-bit distortion and MTM reduced $\chi^2$
within 4\% of unity.
The MTM-corrected mean profiles from these 16 sessions were integrated
to form a polarimetric template profile with a total integration
length of $\sim$100\,hours, shown in \Fig{0437}.
The over-polarization seen from pulse phase 0.88 to 0.94 in this
figure is an instrumental artifact that is currently not understood.
It may be that the unpolarized quantization noise is overestimated by
the two-bit scattered power correction algorithm described in
\App{spc}.
However, a very similar artifact is observed in polarization data
obtained with the second-generation of the Parkes Digital Filter Bank
\citep[PDFB2;][]{ymv+11}, which employs an 8-bit digitizer.
These facts might point to the existence of a non-linear component in
the signal path shared by CPSR2 and PDFB2.
However, the third and fourth generations of the PDFB also regularly
exhibit $\sim5$\% over-polarization when it is not observed using the
latest generation of baseband recording instrumentation at Parkes.
Therefore, it may be only coincidence that a similar artifact is
observed in both PDFB2 and CPSR2 data.


\subsection{Measurement Equation Template Matching}
\label{sec:metm}

Matrix template matching has been previously
utilized to calibrate and derive arrival time estimates from
observations of \psr\ \citep[e.g.][]{vbv+08}.
However, this method cannot be applied directly to \psrtwo\ because
the mean flux of this pulsar at 20~cm \citep[$\sim$6~mJy;][]{mhb+12}
is much lower and the precision of calibrator solutions derived from
similar integration lengths is inadequate.
It is not possible to integrate over longer intervals because the
instrumental response varies with time, primarily due to the
parallactic rotation of the receiver.
Therefore, to calibrate the \psrtwo\ data, a new technique was
developed that combines MTM with MEM.
The new method, called Measured Equation Template Matching (METM),
matches multiple pulsar observations to the template profile, thereby
increasing the precision of the solution.
As with MEM, variations of the instrumental response (e.g. the
parallactic rotation of the receiver) are included in the model.
METM also incorporates measurements of an artificial reference source
(e.g. a noise diode coupled to the receptors), enabling the backend
component of the calibrator solution to be later updated as in
\cite{ovhb04} so that the solution may be applied to other
astronomical sources.
In contrast with MEM, the METM method does not require any additional
observations of a source of known circular polarization to constrain
the boost along the Stokes $V$ axis because the template profile
provides this information.
Using the new METM technique, a detailed history of the instrumental
response at Parkes was derived from the \psr\ data and used to
calibrate the \psrtwo\ data.

\begin{figure*}
\centerline{\includegraphics[angle=0,width=180mm]{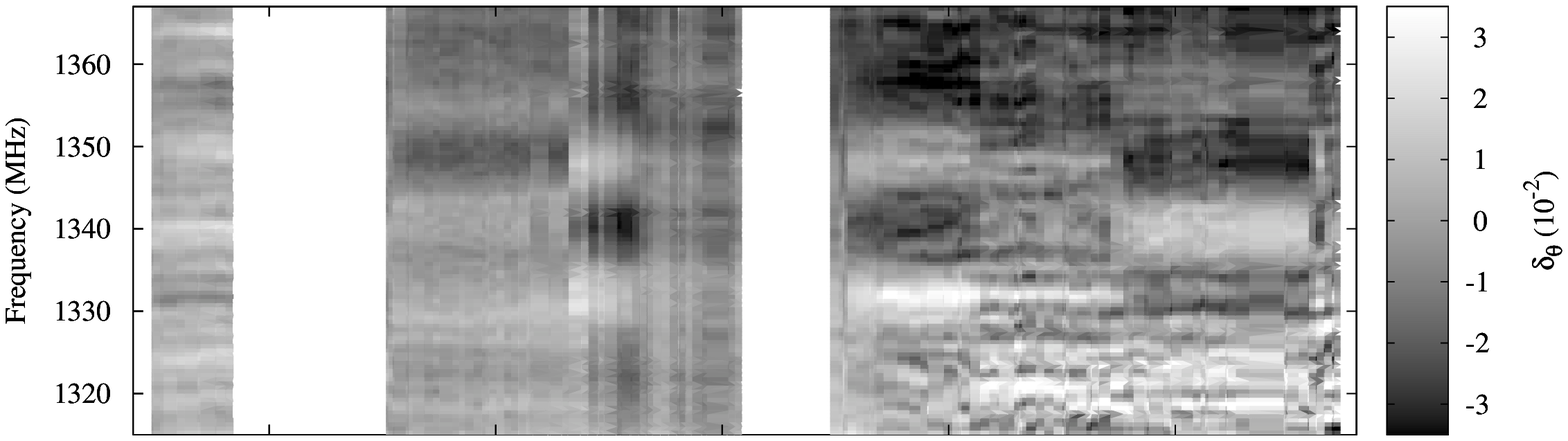}}
\vspace{-2cm}
\centerline{\includegraphics[angle=0,width=180mm]{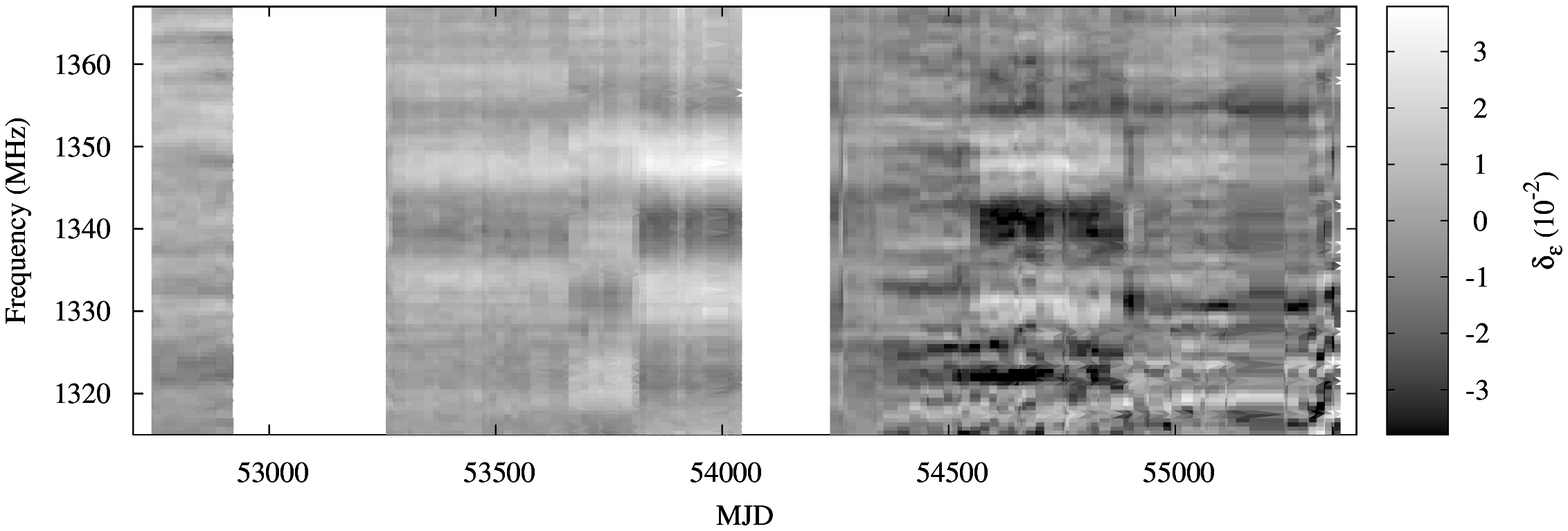}}
\caption{Temporal and spectral variation of receptor cross-coupling. 
  The non-orthogonality parameters, $\delta_\theta$ (upper panel) and
  $\delta_\epsilon$ (lower panel), describe the mixing between total
  intensity and polarized flux in the 21-cm Multibeam receiver and
  down-conversion system of one of the CPSR2 bands.  These estimates
  were derived from regular timing observations of \psr\ using the new
  METM method presented in this paper.  The two gaps in the data
  correspond to periods of maintenance on the 21-cm Multibeam
  receiver.}
\label{fig:coupling}
\end{figure*}

\begin{figure*}
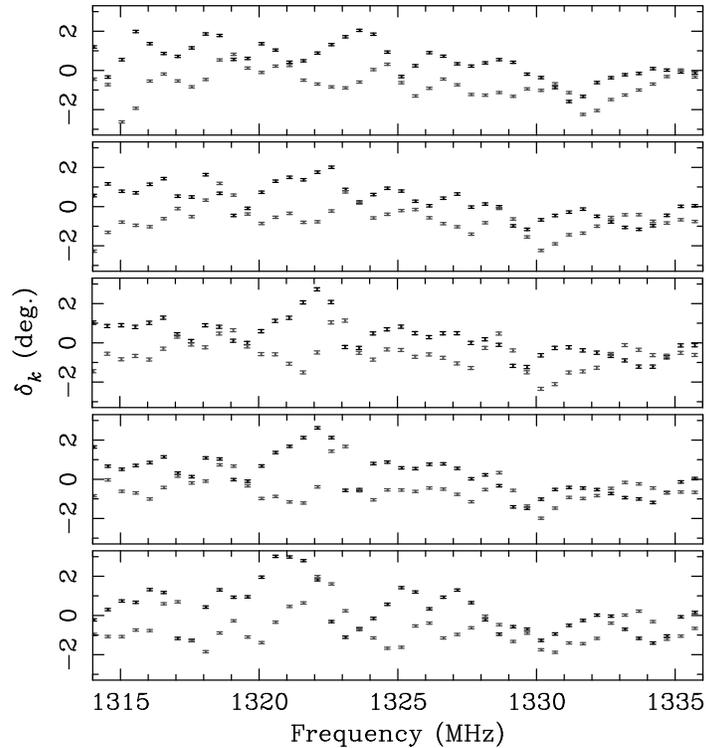

\centerline{\includegraphics[angle=-90,width=86mm]{2010-02-11-0400_1341.eps}}
\centerline{\includegraphics[angle=-90,width=86mm]{2010-02-12-0500_1341.eps}}
\centerline{\includegraphics[angle=-90,width=92.2mm]{2010-02-13-0500_1341.eps}\hspace{6.25mm}}
\centerline{\includegraphics[angle=-90,width=86mm]{2010-02-14-0400_1341.eps}}
\centerline{\includegraphics[angle=-90,width=86mm]{2010-02-16-0400_1341.eps}}
\caption {Temporal and spectral variation of the non-orthogonality
  parameters $\delta_\theta$ (black) and $\delta_\epsilon$ (grey). 
  From top to bottom, the panels plot the solutions obtained on
  11, 12, 13, 14 and 16 February 2010.  The length of each error bar
  is twice the formal standard deviation returned by the least-squares
  fit.}
\label{fig:boost_close}
\end{figure*}

First, the high $S/N$ template profile (created as described in the
previous section) and the new METM method were used to derive a
calibrator solution from every observation of \psr\ with more than 3.5
hours of integration.
A total of $\sim 350$ solutions were produced, spanning
2003 April to 2010 June.  The instrumental response was modeled using
equation (19) of \cite{bri00}, which includes three independent
Lorentz boost transformations described by the differential gain and
the non-orthogonality parameters, $\delta_\epsilon$ and
$\delta_\theta$.
The derived non-orthogonality parameter estimates are plotted in
\Fig{coupling}.
%
%
%
%
%
In addition to smooth variations in $\delta_\theta$ and $\delta_\epsilon$,
there are a number of distinct steps that correspond to physical
changes in the 21-cm Multibeam receiver and downconversion system at Parkes.
The gaps in the data correspond to maintenance periods during which
the Multibeam receiver was unavailable (2003 November 9 to
2004 September 7, and 2006 December 16 to 2007 May 17).
Immediately following the first maintenance period, the receptor
cross-coupling remains stable for a period of approximately one year.
The distinct changes in state that occur around MJD 53660 (2005
October 17) and MJD 53800 (2006 March 6) are due to
changes in the downconversion system (the signal paths for the two
CPSR2 bands were swapped).  
Following the second maintenance period, the cross-coupling parameters
are notably less stable in both time and frequency.
%
%
\Fig{boost_close} provides a closer look at the variations of these
parameters over a period of one week and a bandwidth of 20~MHz.
In addition to significant quasi-periodic variations in these
parameters as a function of radio frequency, the spectral structure
appears to drift as a function of time.  For example, from 11 to 12
February 2010, the main features in the spectrum have
shifted toward lower frequencies by $\sim1.5$~MHz.
After MJD $\sim$ 54300, it is invalid to assume that the receptor
cross-coupling remains stable for any period of time longer than a
day.

\Fig{coupling} plots the non-orthogonality parameters in units
of centiradians; these values can be multiplied by the estimates of
$\tau_\beta$ listed in \Tab{theory} to approximate the variation in
systematic arrival time error introduced by receptor
cross-coupling.  
For example, for \psrtwo, the peak-to-peak variations of $\sim 0.06$
radians in both $\delta_\theta$ and $\delta_\epsilon$ are predicted to induce
systematic timing errors of the order of 1.7\,$\mu$s.
However, after integrating over all frequency channels, the net effect
of the cross-coupling variations will be diminished.
Furthermore, receptor cross-coupling describes only two of the three
forms of instrumental distortion.
In addition to the presumption of zero receptor cross-coupling,
calibration techniques based on the ideal feed assumption typically
also incorporate the false premise that the artificial reference
source illuminates both receptors identically.
Under this assumption, any intrinsic imbalance in the induced
amplitude of the reference source signal in each receptor will be
misinterpreted as differential gain.
In contrast, the METM technique includes direct estimation of the
Stokes parameters of the reference source
\citep[as in Fig.~2 of][]{van04c}.
Using these estimates, the systematic error in differential gain
$\delta_\gamma$ due to unbalanced reference source amplitudes
may be derived.

\begin{figure*}
\centerline{\includegraphics[angle=0,width=150mm]{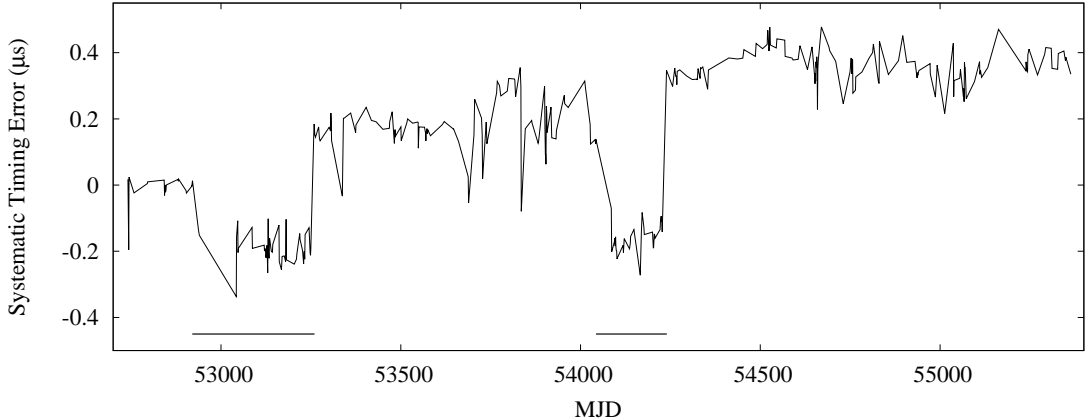}}
\caption {Systematic arrival time error induced by instrumental distortion
  in the 21-cm Multibeam and H-OH receivers at Parkes. The periods during
  which the H-OH receiver was used are marked by two horizontal lines
  near the MJD axis. }
\label{fig:systematic_error}
\end{figure*}

Given estimates of $\delta_\gamma$, $\delta_\theta$ and
$\delta_\epsilon$ as a function of frequency, it is possible to
directly compute the systematic arrival time error at each epoch by
applying the instrumental distortion transformation to a copy of the
template profile and estimating the induced phase shift between the
template and its distorted copy.
Application of the METM-derived instrumental distortion parameters to
the \psrtwo\ template profile yields the systematic
timing error shown in \Fig{systematic_error}.
The instrumental distortion transformation is applied independently in
each frequency channel before integrating over frequency.
The induced phase shift is computed using only the total intensity
profile and the Fourier-domain template matching algorithm described by
\citet{tay92}.
During the periods of maintenance on the Parkes 21-cm Multibeam
receiver, the solutions derived from observations made with the \mbox{H-OH}
receiver have been used.
Uncalibrated instrumental distortion introduces peak-to-peak
variations in systematic timing error of the order of 600\,ns;
the largest jumps in timing error occur when switching between
receivers.
These systematic errors are present in arrival times derived from
either uncalibrated or inaccurately-calibrated \psrtwo\ observations.
As shown in the following section, these errors are significantly
reduced after calibration with the METM-derived solutions.


\subsection{Arrival time estimation and analysis}
\label{sec:pat}

\begin{figure*}
\centerline{\includegraphics[angle=-90,width=86mm]{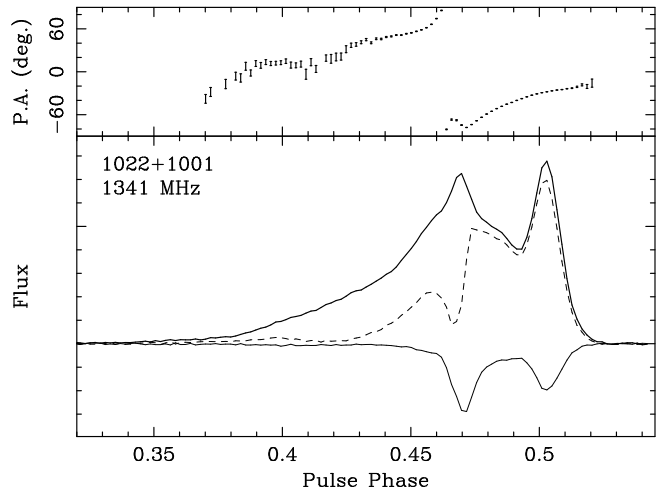}}
\caption {Mean polarization of \psrtwo, plotted as a function of pulse
  phase using cylindrical coordinates: position angle (top panel),
  linearly polarized intensity, $L=\sqrt{Q^2+U^2}$ (dashed line),
  circular polarization, $V$ (thin solid line) and total intensity
  (thick solid line).  Data were integrated over a 64\,MHz band
  centered at 1341\,MHz for approximately 67~hours. }
\label{fig:1022}
\end{figure*}

The solutions produced using the new METM method were applied to
calibrate the 5-minute \psrtwo\ integrations as in \cite{ovhb04}.
The calibrated data were then integrated in frequency and time to form
approximately 64-minute integrations.
The observations with the greatest $S/N$ were added to form a
full-polarization template profile with an integration length of
$\sim67$ hours, shown in \Fig{1022}.

To enable a quantitative comparison between the analysis described in
this paper and previous work, the \psrtwo\ observations were also
calibrated using the ideal feed assumption employed by \cite{hbo04}.
Arrival times for the two datasets were then derived using only the total
intensity profile \citep{tay92} and matrix template matching \citep{van06}. 
These variations produced four sets of arrival time estimates,
calibrated using either full measurement equation template matching (METM) or
the ideal feed assumption (IFA) and timed using either matrix
template matching (MTM) or standard total intensity (STI) methods.
The post-fit arrival time residuals for these four sets are compared
in \Tab{residual}.
\begin{table*}
\caption{Arrival Time Residual Standard Deviation and Merit} 
\begin{center} 
\begin{tabular}{ccc} 
\tableline 
\tableline 
Method & $\sigma_\tau$ ($\mu$s) & $\chi^2/N_\mathrm{free}$ \\
\tableline 

IFA--STI & 1.9  & 2.2 \\
METM--STI & 1.4  & 2.1 \\

IFA--MTM & 0.92 & 1.1 \\
METM--MTM & 0.88 & 1.0 \\

\tableline 
\end{tabular} 
\end{center} 
\label{tab:residual}
\end{table*}

After calibration based on the ideal feed assumption (IFA) and arrival
time estimation with the standard method of template matching
using only the total intensity profile (STI), the residuals are of
similar quality to those presented by \citet{hbo06}.
That is, compared to uncalibrated data, the IFA-calibrated data are
equally adversely affected by distortions to the total intensity
profile.
This is not surprising, given that the IFA does not apply to the 21-cm
Multibeam receiver at Parkes \citep{van04c}.
Calibration via measurement equation template matching (METM) reduces
both the standard deviation of the arrival time residuals and the
reduced $\chi^2$ of the model fit.
Assuming that the systematic error removed by METM is uncorrelated
with the remaining noise in the METM--STI residuals, 
instrumental distortion of the IFA-calibrated total intensity profile
gives rise to timing errors with a standard deviation of approximately
1.3~$\mu$s.

For both METM- and IFA-calibrated data sets, matrix template matching
(MTM) yields arrival times of similar quality.
This indicates that MTM is able to correct the remaining instrumental
calibration errors in the IFA data and produce arrival time estimates
with reduced systematic error.
It is not possible to test MTM on uncalibrated data because the
variation of instrumental response with frequency leads to bandwidth
depolarization, which cannot be modeled using a single Jones matrix.
That is, although inaccurate, calibration based on the IFA may be
sufficient to avoid depolarization.

Focusing on only the METM-calibrated data, comparison of arrival times
generated using STI and MTM demonstrates that MTM reduces the standard
deviation of arrival time residuals by approximately 35\%, which is
consistent with the relative uncertainty predicted in \Tab{theory} and
by \citet{van06}.
When compared to the results of the typical data analysis employed in
most high-precision timing experiments to date (IFA--STI), matrix
template matching reduces arrival time residuals by over 50\%.
The corresponding improvement in the experimental sensitivity of this
dataset is equivalent to that of quadrupling the integration length
of each observation.

\subsection{Results}
\label{sec:results}

\begin{figure*}
\centerline{\includegraphics[angle=0,width=86mm]{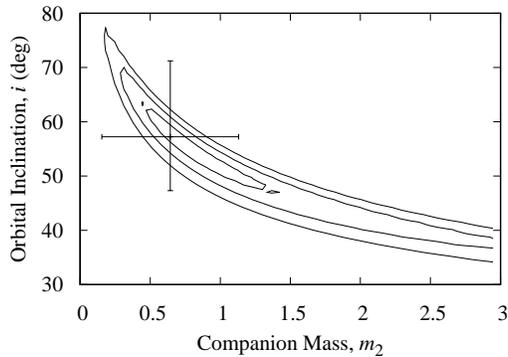}}
\caption {Map of $\Delta\chi^2$ as a function of binary companion mass
  and orbital inclination angle.  From innermost to outermost, the
  contours represent $\Delta\chi^2$ = 1, 4, and 9.  The error bars
  indicate the best-fit solution and the formal errors (one standard
  deviation) returned by {\sc tempo2} for the shape and range of the
  Shapiro delay.}
\label{fig:m2sini}
\end{figure*}

The results presented in this section are based on the data calibrated
using METM and arrival time estimates derived using MTM.
\Tab{parameters} lists the best-fit timing model parameters 
obtained using the {\sc tempo2} analysis software \citep{hem06,ehm06}.
\begin{table*}
\begin{center} 
\caption{Best-fit model parameters for \psrtwo}
\begin{tabular}{lr}
 \tableline \tableline
Parameter Name & Value \\
\tableline
Reference clock & TT(TAI) \\
Planetary Ephemeris & DE414 \\
$P$ epoch (MJD) & 53869.00016629324189 \\
Pulse period, $P$ (ms) & 16.4529299518323(2) \\
Period derivative, $\dot{P}$ (10$^{-20}$) & 4.33399(6) \\
Ecliptic longitude, $\lambda$ (\degr) & 153.865881(3) \\
Ecliptic latitude, $\beta$ (\degr) & $-$0.06(4) \\
Proper motion in $\lambda$ (mas yr$^{-1}$) & $-$15.97(2) \\
Proper motion in $\beta$ (mas yr$^{-1}$) & 38(19) \\
Annual parallax, $\pi$ (mas) & 1.4(2) \\
Binary Model & DDH \\
Orbital period, $P_{\rm b}$ (days) & 7.805135(1) \\
Projected semi-major axis, $x$ (lt-s) & 16.76541(4) \\
Orbital eccentricity, $e$ ($10^{-5}$) & 9.702(7) \\
Epoch of periastron, $T_0$ (MJD) & 53876.1038(2) \\
Longitude of periastron, $\omega$ (\degr) & 97.757(9) \\
Orthometric amplitude, $h_3$ ($\mu$s) & 0.52(7) \\
Orthometric ratio, $\varsigma$ & 0.5(1) \\
Periastron advance, $\dot \omega$ (\degr yr$^{-1}$) & 0.009(3) \\
Secular variation of $x$, $\dot x$ ($10^{-15}$) & 14(1) \\
\tableline
\end{tabular}
\end{center} 
\label{tab:parameters}
\tablecomments{For each parameter, the formal uncertainty (one standard
  deviation) in the last digit quoted is given in parentheses.} 
\end{table*} 
These results include a significant detection of the Shapiro delay;
after subtracting two degrees of freedom from the least-squares fit
with the addition of the shape $s=\sin i$ and range $r=m_2$
parameters, $\chi^2$ is reduced by $\sim20$\% (from $\sim330$ to
$\sim260$).
%
%
%
Although the detection is significant, the elongated contours of
constant $\Delta\chi^2$ shown in \Fig{m2sini} demonstrate that $s$ and
$r$ are highly covariant; 
the coefficient of correlation between these two model parameters is
$-$0.99.
Furthermore, the constraint on the companion mass is
not very informative;
the 95\% confidence interval given by the projection of the
$\Delta\chi^2$ = 4 contour onto the $m_2$ axis ranges
from 0.25 to $>3$ solar masses.


The shape and range parameters are highly covariant because, in nearly
circular orbits that are only moderately inclined with respect to the
line of sight,
the Shapiro delay is readily absorbed in the Roemer delay by
modification of the Keplerian orbital parameters.
The unabsorbed remnant of the Shapiro delay is best described as a
weighted sum of harmonics of the orbital frequency using the
orthometric parameterization introduced by \cite{fw10}.
Accordingly, \Tab{parameters} lists the best-fit values of the ratio of the
amplitudes of successive harmonics
\[
\varsigma = { \sin i \over 1 + |\cos i| }
\]
and the amplitude of the third harmonic
\[
h_3 = m_2 \varsigma^3.
\]
The values of $s$ and $r$ derived from $\varsigma$ and $h_3$ are
nearly identical to the values yielded by directly modeling the shape
and range of the Shapiro delay.
However, the coefficient of correlation between $h_3$ and $\varsigma$
is only $-$0.38.
There is close agreement between the orbital inclination angle, $i
\sim 57\degr$, constrained by the Shapiro delay and previous estimates
of the inclination of the pulsar spin axis, $\zeta \sim 61\degr$
\citep{xkj+98} and $\zeta \sim 55 \degr$ \citep{rk03}, constrained
using the rotating vector model \citep{rc69a,ew01}.
This agreement is compatible with the expectation that the pulsar
spin and orbital angular momentum vectors are aligned during the
period of mass accretion that led to the formation of the millisecond
pulsar \citep[e.g.][]{tc99}.

The best-fit model parameters also include the first significant
detections of the precession of periastron $\dot\omega$ and the
secular variation of the projected semi-major axis of the orbit,
$\dot x$.
%
%
%
The estimate of
$\dot\omega\sim(9\degr\pm3\degr)\times10^{-3}$yr$^{-1}$ is consistent
with the value predicted by General Relativity, $\dot\omega^{\rm
  GR}\sim8\degr\times10^{-3}$yr$^{-1}$, based on the masses of the
pulsar and its companion yielded by the Shapiro delay detection.
The measured value of $\dot x$ is seven orders of magnitude larger
than the general relativistic prediction, $\dot x^\mathrm{GR} \sim 5
\times 10^{-21}$, because it is dominated by the secular variation
due to proper motion \citep{kop96}.
This geometric effect can be used to place an upper limit on the orbital
inclination angle; i.e.
%
$\tan i < \mu { x / \dot x }$,
%
where $\mu=42 \pm 18$ mas yr$^{-1}$ is the total proper motion.
The resulting constraint, $i < 83\degr$, rules out only 12\% of
possible orientations and is consistent with the stronger constraint
yielded by measurement of the Shapiro delay.

The proper motion of the system also induces a quadratic Doppler shift
\citep{shk70} that contributes $T \mu^2 d / c$ to observed period
derivatives, where $T$ is the period (e.g.\ spin or orbital), $d$ is
the distance to the pulsar, and $c$ is the speed of light.
The magnitude of the Shklovskii effect is estimated using the distance
$d_\pi = 750^{+100}_{-90}$ parsec derived from the annual
trigonometric parallax after correction for Lutz-Kelker bias, $\pi =
1.340^{+0.175}_{-0.165}$ mas \citep{vlm10}.
The Shklovskii contribution to the spin period derivative ($\sim
4.9 \times 10^{-20}$) is slightly larger than its measured value,
which would imply that the intrinsic $\dot P$ of the pulsar is
negligible.
However, the predicted contribution to the orbital period derivative
($\sim 2 \times 10^{-12}$) is around 20 times larger than the upper
limit derived by including $\dot P_\mathrm{b}$ in the timing model.
The relative uncertainties in the above estimates of the Shklovskii
effect are almost 100\%, primarily due to the large uncertainty in the
estimate of the proper motion in ecliptic latitude.
Using only the proper motion in ecliptic longitude,
the predicted kinematic contribution to $\dot P_\mathrm{b}$ is
only three times larger than the timing-derived upper limit.

To explain the remaining discrepancy, three other contributions to the
observed $\dot P_\mathrm{b}$ are considered \citep[as in][]{dt91}.
First, the loss of energy due to gravitational radiation as predicted
by General Relativity contributes around $-3 \times 10^{-16}$,
approximately 4 orders of magnitude smaller than the Shklovskii
effect.
Second, the contribution due to differential rotation about the
Galactic centre (approximately $-7 \times 10^{-15}$) can also be
neglected.
Finally, the acceleration of the \psrtwo\ system in the Galactic
gravitational potential is considered.
Using the parallax distance $d_\pi$ and the Galactic latitude
$b=51\degr$, the component of gravitational
acceleration perpendicular to the Galactic disk, $K_z \sim
6.6\times10^{-11}$~m~s$^{-2}$ \citep{bah84}
contributes approximately $-1.2 \times 10^{-13}$ to the observed $\dot
P_\mathrm{b}$.
At large heights above the Galactic disk, the relative uncertainty in
$K_z$ is more than a factor of two; therefore, this contribution is
large enough to negate the Shklovskii effect, resulting in an observed
$\dot P_\mathrm{b}$ that is consistent with zero.
%


\section{Discussion}
\label{sec:discussion}

Instrumental distortion of the total intensity profile introduces
significant systematic timing errors that are correlated between
pulsars observed with the same instrument.
Therefore, high-fidelity polarimetry is inextricably linked with the
long-term objectives of high-precision timing and the primary goals of
pulsar timing array experiments.
This paper presents a novel method of calibrating the instrumental
response by exploiting the long-term stability of the polarized
emission from a millisecond pulsar.
Full-polarization timing observations of \psr\ are used to model
variations in the 21-cm Multibeam and H-OH receivers and
down-conversion system at Parkes.
Over a period of $\sim7.2$ years, temporal and spectral variations of
the receptor cross-coupling introduces systematic error of the order
of 1\,$\mu$s in arrival time estimates derived from observations of
\psrtwo.
High-fidelity calibration of instrumental polarization and arrival
time estimation using matrix template matching is demonstrated to
correct systematic error and double the sensitivity of the experiment,
yielding significant detections of the Shapiro delay, the precession
of periastron, and the secular variation of the projected semi-major
axis of the orbit.
The improvements in both timing accuracy and precision are consistent
with the hypothesis that the average polarized emission from these
millisecond pulsars is stable over the timescales of relevance to this
experiment.

With a modest instrumental bandwidth of 64~MHz and median integration
length of 64~minutes, \psrtwo\ yields arrival time residuals with an
uncertainty-weighted standard deviation of only $\sigma_\tau$ =
880\,ns, roughly five orders of magnitude smaller
than the pulsar spin period $P \sim 16.5$~s.
The remarkable timing precision of these observations supports the
conclusions of \cite{hbo04}, who argue that the \psrtwo\ profile
variations observed by \cite{kxc+99} and \cite{rk03} are not intrinsic
to the pulsar and are most likely due to instrumental calibration
errors.

The precision of the arrival time data presented in this
paper places \psrtwo\ among the top ten sources regularly observed as
part of the Parkes Pulsar Timing Array (PPTA) project.
%
%
Other PPTA sources may also benefit from the application of METM;
e.g. six of the sources listed in \Tab{theory} have instrumental
distortion susceptibility factors ranging from 200 ns to 340 ns.
Referring to the measured values of the non-orthogonality parameters
plotted in \Fig{coupling}, these factors roughly correspond to 
systematic timing error variations of the order of 1 $\mu$s.
Therefore, the timing precision of these pulsars would likely be improved
by application of the methods developed for this study.
For PSR~J1744$-$1134 and PSR~J2129$-$5721, matrix template matching is
predicted to yield arrival times with greater uncertainty than those
derived from the total intensity profile.
%
%
In these two cases, the best results would be obtained by calibrating
using METM and deriving arrival times using STI.

In a wider analysis of all of the pulsars in the PPTA, \citet{mhb+12}
calibrated approximately 5.9~years of \psrtwo\ timing observations using
MEM-derived solutions.
Using typical integration lengths of 64~min and an instrumental bandwidth of
256~MHz, this study yielded arrival time residuals with an
uncertainty-weighted standard deviation of 1.7~$\mu$s and reduced
$\chi^2\sim9.3$.
The residual noise is approximately 2.4 times greater than expected
based on simple extrapolation of the METM--STI results presented in
\Tab{residual}.
This discrepancy may be explained by the fact that the PDFB
instruments have a non-linear response that introduces
over-polarization, which cannot be calibrated using MEM.
The accuracy of MEM-based calibration is also limited by the fact that
there is no unique solution to the measurement equation when the only
constraining transformation is the geometric rotation of the receptors
with respect to the sky \citep{van04c}.
To constrain the otherwise degenerate boost along the Stokes $V$ axis,
it is necessary to include observations of a source that is assumed to
have negligible circular polarization
(e.g., for the MEM fits performed in \Sec{template}, observations of
Hydra~A were used to constrain $\delta_\epsilon$).
However, the use of an unpulsed source of radiation as a constraint
is intrinsically susceptible to variability in other contributions to
the system temperature, including receiver noise and ground spillover.
That is, it is safer to assume that the polarized emission from \psr\
is constant than it is to rely on a source of unpulsed radiation
during calibration.
The results presented by \citet{mhb+12} may also be limited by
the instability of the receptor cross-coupling, as shown in
\Fig{coupling}.  The temporal variations of the
non-orthogonality parameters are sufficiently resolved only via
application of the METM method presented in this paper, which
yields a seven-fold increase in the number of available calibrator
solutions.

As noted in previous studies
\citep{kxc+99,rk03,hbo04,mhb+12}, other phenomena may also currently
limit the timing precision of \psrtwo.
For example, owing to its low ecliptic latitude, the radio
signal from this pulsar is subject to significant dispersive delays in
the solar wind \citep{yhc+07a,ychm12}.
No corrections for dispersion variations were applied to the data
presented in this work; however, observations made when the line of
sight to \psrtwo\ passes near the Sun (around late August of each
year) were excluded from the data set.
In addition to fluctuations in dispersion, the observed flux of the
pulsar varies as a function of time and radio frequency due to
scintillation in the interstellar medium.
The average profile of \psrtwo\ varies significantly as a
function of radio frequency \citep[e.g. see Figure 1 of][]{rk03} and,
when modulated by interstellar scintillation, the
frequency-integrated mean profile may fluctuate with time.
The potentially significant arrival time estimation errors induced by
this effect could be mitigated through the use of a
frequency-dependent template profile.
%
%

It is reasonable to expect that \psrtwo\ timing will be improved
by the current generation of instrumentation at the Parkes Observatory.
The data presented in this paper were observed using a system with low
dynamic range that performed analog-to-digital conversion of the radio
signal with only two bits per sample.
Two-bit quantization is an intrinsically non-linear process and the
techniques applied to restore linearity (dynamic output level setting)
and mitigate quantization noise (scattered power correction) are based
on a linear approximation to the response of the digitizer \citep{ja98}.
As discussed in \App{spc}, the scattered power correction
algorithm applied in this analysis is accurate only to first order and
it is possible that overestimation of the unpolarized scattered power
contributes to the over-polarization noted in \Fig{0437}.
To overcome the limitations of two-bit sampling, a new baseband
recording and processing system with greater dynamic range was
commissioned at the Parkes Observatory in 2010 April.
Designed in collaboration with the Center for Astronomy Signal
Processing and Electronics Research (CASPER) at Berkeley, the
CASPER-Parkes-Swinburne Recorder (CASPSR) digitizes a
dual-polarization 400~MHz band with eight bits per sample and performs
real-time radio frequency interference excision based on the spectral
kurtosis estimator \citep{ng10}.
This system currently operates in parallel with the third
and fourth generations of the Parkes Digital Filter Bank (PDFB)
instruments as part of the PPTA program.

The calibration techniques applied in this paper could potentially
be improved by reducing the number of degrees of freedom in the
estimates of the receptor cross-coupling parameters.
For example, this could be achieved by fitting an analytic model to
the temporal and spectral variations of the calibrator solutions or by
employing a lossy compression algorithm that can be applied to
irregularly sampled data.
The smoothing effected by such a transformation would reduce the
instantaneous noise in the applied calibrator solutions and might
also provide a robust means of interpolating between solutions.
However, further refinements of the calibration technique may yield
only marginal improvements to arrival time accuracy and precision.
\Tab{residual} demonstrates that, even when the data are calibrated
using the inaccurate ideal-feed assumption, matrix template matching (MTM)
yields arrival time estimates that are nearly as good as those
derived from data calibrated using the new METM technique described
in this paper.
This seems to suggest that, as long as the signal is not depolarized
by integrating over time and/or frequency, the fidelity of the
calibration technique has negligible impact on arrival times derived
using MTM.

This conclusion may have a significant impact on the design of the next
generation of instrumentation for high-precision timing.
For example, it has been asserted that in order to achieve timing
precision of the order of 100~ns, the Square Kilometre Array (SKA)
must achieve net polarization purity corresponding to $\beta<10^{-4}$
\citep{ckl+04}.
However, even when non-orthogonality as large as $\beta\sim10^{-2}$ is
ignored, as is the case when the 21-cm Multibeam receiver at Parkes is
assumed to be ideal, MTM yields accurate arrival time estimates.
That is, by relaxing the requirement for polarization purity, the
application of MTM has the potential to reduce the cost of SKA
development for high-precision timing.

All of the software required to perform measurement equation modeling,
polarimetric calibration, and matrix template matching is freely
available as part of {\sc psrchive}~\citep{hvm04}; the use of this
software is demonstrated by \citet{vdo12} and more fully documented
online\footnote{http://psrchive.sourceforge.net/manuals}.

\acknowledgements

I am grateful to Matthew Bailes, Paul Demorest, Mike Keith, Michael
Kramer, Stefan Os{\l}owski, and John Reynolds for helpful discussions
during this research project and for insightful comments that greatly
improved this report.
Joris Verbiest implemented and assisted with {\sc tempo2} support of
ecliptic coordinates and the orthometric parameterization of the
Shapiro delay.
The Parkes Observatory is part of the Australia Telescope which is
funded by the Commonwealth of Australia for operation as a National
Facility managed by CSIRO.

\appendix

\section{Probability Distribution of the Polarimetric Invariant}
\label{app:invariant}

\begin{figure*}
\centerline{\includegraphics[angle=-90,width=86mm]{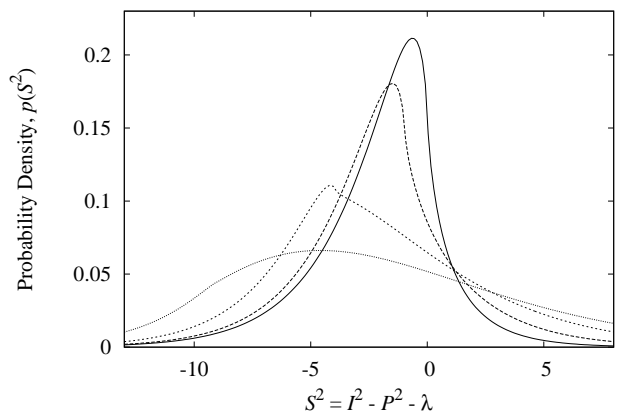}}
\caption {Probability density of the centralized polarimetric
  invariant, $S^2=I^2-P^2-\lambda$, where $I$ is the total intensity,
  $P$ is the polarized flux, and $\lambda = \langle I \rangle^2 -
  \langle P \rangle^2$ is the noncentrality of $S^2$.  The standard
  deviation in each of the four Stokes parameters is unity; therefore,
  the mean value of $S^2$ is $-2$.  The four curves correspond
  to $\langle I \rangle$ = 0 (solid), 1, (long dash), 2 (short
  dashed), and 3 (dotted).  In each case, $\langle P \rangle$ = 0.
  For large values of $\langle I \rangle$, the probability density of
  $S^2$ approaches a normal distribution. }
\label{fig:invint}
\end{figure*}

The invariant profile is formed by computing the Lorentz interval
$\Sinv$ as a function of pulse longitude, where
$\Sinv^2\equiv S_0^2-|\mbf{S}|^2$.  
Assuming that the noise in each of the Stokes parameters is
independent and normally distributed with standard deviation
$\varsigma$, define the normalized Stokes parameters
$S^\prime_k=S_k/\varsigma$.  The noise in $S^{\prime2}_0$ has a
noncentral $\chi^2$ distribution with one degree of freedom and
noncentrality parameter $\lambda=S^{\prime2}_0$.  Likewise, the noise
in $|\mbf{S}^\prime|^2$ has a noncentral $\chi^2$ distribution with three
degrees of freedom and $\lambda=|\mbf{S}^\prime|^2$.  The distribution
of the noise in $\Sinv^2$ is given by the cross-correlation of the
distributions of $S_0^{\prime2}$ and $|\mbf{S}^\prime|^2$; examples
are shown in \Fig{invint}.
The variance of $\Sinv^2$ is equal to $4\norm{S^\prime}^2 + 8$, where
$\norm{S}$ is the Euclidean norm of the Stokes four-vector, defined by
$\norm{S}^2\equiv S_0^2+|\mbf{S}|^2$.
As all four Stokes parameters vary as a function of pulse longitude,
the noise in $\Sinv^{\prime2}$ is heteroscedastic,
which violates one of the basic premises of
most template-matching algorithms.

Both the distribution of the invariant interval and its tendency toward zero
for highly polarized radiation adversely affect its usefulness in
template matching (as a replacement for the total intensity) and as a
normalization factor during measurement equation modeling
\citep[MEM;][]{van04c}.
Gain normalization is necessitated by interstellar scintillation,
which causes the received flux of the pulsar to vary as a function of
time and radio frequency.
To address the problems with the invariant interval during the application
of MEM in \Sec{template}, the Stokes
parameters were normalized by the sum of the invariant interval over all
on-pulse longitudes.
For the typical observation of \psr, this summation reduces the
relative error of the normalization factor by $\sim97\%$, thereby
yielding more normally distributed normalized Stokes parameters.

\section{Folded Profile Scattered Power Correction}
\label{app:spc}

The data presented in this paper were processed using a new scattered
power correction algorithm that can be used to correct digitization
distortions to folded pulsar profiles.  In the case of two-bit sampled
voltages, the scattered power (or quantization noise) is $\sim 12$\%
of the total power.
The derivation of the method starts with the mean digitized power
$\hat\sigma^2$ given by equation (A5) of \citet[hereafter JA98]{ja98},
\begin{equation}
\hat\sigma^2 = \sum_\Phi {\mathcal P}(\Phi)f(\Phi),
\label{eqn:A5}
\end{equation}
where $\Phi$ is the fraction of samples that fall between the chosen
thresholds, $f(\Phi)$ is the digitized power as a function of $\Phi$,
given by equation (A4) of JA98, and ${\mathcal P}(\Phi)$ is the
discrete probability distribution for $\Phi$, given by equation (A6)
of JA98.
The parameter $\Phi$ is eliminated by the summation in the above equation;
however, both $f(\Phi)$ and ${\mathcal P}(\Phi)$ are also
parameterized by the expectation value $\langle\Phi\rangle$.
Therefore, \eqn{A5} represents the relationship between the mean
digitized power and the mean value of $\Phi$.
That is, given the mean digitized power $\hat\sigma^2$, \eqn{A5} can
be inverted to compute $\langle\Phi\rangle$, which can in turn be used
to estimate the mean undigitized power $\sigma^2$ and the mean
scattered power $1-A$ via equations (45) and (43) of JA98,
respectively.

Equation~(A5) of JA98 can be inverted using the Newton-Raphson method
and the partial derivatives of equations (A4) through (A6) of JA98
with respect to $\langle\Phi\rangle$.
These are simplified in the case of two-bit sampling by noting that
equation (A4) of JA98 reduces to
\[
f(\Phi) = \langle\Phi\rangle y_3^2(\Phi) + (1-\langle\Phi\rangle) y_4^2(\Phi)
\]
where $y_3(\Phi)$ and $y_4(\Phi)$ are given by equations (41)
and (40) of JA98.

The folded profile scattered power correction algorithm is based on
the following assumptions and approximations.
First, at the time of recording the astronomical signal, it is
necessary that the baseband voltages are sampled using the optimum
two-bit input thresholds defined in Table 1 of JA98.
To first order, this condition is satisfied by excluding data with
excessive two-bit distortion as defined by \eqn{d2bit} of this paper.
Second, equations (45) and (43) of JA98 only approximately relate the
expectation values of $\Phi$, $\sigma^2$ and $A$.
For example, the relationship between $A$ and $\sigma^2$ defined by
equation (43) of JA98 is concave down (see Figure 3 of JA98);
therefore, Jensen's inequality dictates that the value of $A$
estimated from the mean undigitized power will be greater than the
expectation value of $A$.
Consequently, the mean fractional scattered power may be
systematically underestimated.
This limitation cannot be overcome without prior knowledge of the
distribution of total intensity fluctuations intrinsic to the pulsar
signal.
Finally, it is assumed that the signal in each frequency channel has
not been significantly altered, such that there is a well-defined
relationship between the mean undigitized power and the mean digitized
power.
For example, after phase-coherent dedispersion removal, the flux
density over a given time interval no longer represents the voltage
fluctuations in the digitizer over that interval; the previously
smeared pulsar signal will be recovered and the scattered power will
be smeared.
To estimate the scattered power using coherently-dedispersed digitized
power, the dispersion smearing in each frequency channel must be less
than or of the order of the time resolution of the folded profile.
This condition is satisfied in the 20-cm timing observations of \psr\
and \psrtwo\ presented in this paper.


\end{document}